\documentclass[sigconf]{acmart}

\copyrightyear{2026}
\acmYear{2026}
\setcopyright{cc}
\setcctype{by}
\acmConference[CHI EA '26]{Extended Abstracts of the 2026 CHI Conference on Human Factors in Computing Systems}{April 13--17, 2026}{Barcelona, Spain}
\acmBooktitle{Extended Abstracts of the 2026 CHI Conference on Human Factors in Computing Systems (CHI EA '26), April 13--17, 2026, Barcelona, Spain}
\acmDOI{10.1145/3772363.3798567}
\acmISBN{979-8-4007-2281-3/2026/04}

\usepackage{graphicx}  
\usepackage{subcaption} 
\usepackage{csquotes}
\usepackage{siunitx}
\usepackage{enumitem}
\begin{document}

\title{Who Explains Privacy Policies to Me? Embodied and Textual LLM-Powered Privacy Assistants in Virtual Reality}

\author{Vincent Freiberger}
\orcid{0009-0000-8549-9119}
\affiliation{%
  \institution{Center for Scalable Data Analytics and Artificial Intelligence (ScaDS.AI) Dresden/Leipzig, Leipzig University}
  \city{Leipzig}
  \country{Germany}
}
\email{freiberger@cs.uni-leipzig.de}

\author{Moritz Dresch}
\orcid{0009-0006-3639-3864}
\affiliation{
    \institution{LMU Munich}
    \city{Munich}
    \country{Germany}
}
\email{mo.dresch@campus.lmu.de}

\author{Florian Alt}
\orcid{0000-0001-8354-2195}
\affiliation{
    \institution{LMU Munich}
    \city{Munich}
    \country{Germany}
    }
\affiliation{
    \institution{University of the Bundeswehr Munich}
    \city{Munich}
    \country{Germany}
    }
\email{florian.alt@ifi.lmu.de}

\author{Arthur Fleig}
\authornote{Equal last author contribution.}
\orcid{0000-0003-4987-7308} 
\affiliation{%
  \institution{Center for Scalable Data Analytics and Artificial Intelligence (ScaDS.AI) Dresden/Leipzig, Leipzig University}
  \city{Leipzig}
  \country{Germany}
}
\email{arthur.fleig@uni-leipzig.de}

\author{Viktorija Paneva}
\authornotemark[1]
\orcid{0000-0002-5152-3077}
\affiliation{
    \institution{LMU Munich}
    \city{Munich}
    \country{Germany}
    }
\email{viktorija.paneva@ifi.lmu.de}

\begin{abstract}

Virtual Reality (VR) systems collect fine-grained behavioral and biometric data, yet privacy policies are rarely read or understood due to their complex language, length, and poor integration into users' interaction workflows.
To lower the barrier to informed consent at the point of choice,
we explore a Large Language Model (LLM)-powered privacy assistant embedded into a VR app store to support privacy-aware app selection.
The assistant is realized in two interaction modes: a text-based chat interface and an embodied virtual avatar providing spoken explanations. 
We report on an exploratory within-subjects study $(N = 21)$ in which participants browsed VR productivity applications under unassisted and assisted conditions.
Our findings suggest that both interaction modes support more deliberate engagement with privacy information and decision-making, with privacy scores primarily functioning as a veto mechanism rather than a primary selection driver.
The impact of embodied interaction varied between participants, 
while textual interaction supported reflective review.

\end{abstract}



\maketitle

\section{Introduction}
\label{sec:intro}

VR systems rely on a range of high-fidelity sensing technologies, such as cameras, eye-tracking, and inertial measurement units, to capture user behavior with millimeter-level precision~\cite{eyeTracking,viveTracking,XR}. While these capabilities enable immersive experiences, they also transform subtle movements and gaze patterns into highly identifiable biometric data. 
As little as 100 seconds of motion data can identify individuals within a pool of 55,541 with 94.33\% accuracy~\cite{motionIdentification50k}. 
VR data can reveal potentially sensitive inferences such as cognitive load, emotional states, and even sexual orientation~\cite{motionIdentification}, creating a form of biometric psychography~\cite{heller2020watching} more revealing than conventional web browsing.
Despite these risks, VR privacy documentation remains static and opaque, often buried in external 2D menus that users rarely consult before installation~\cite{Prillard2024EthicalDF,Zhan2024VPVetVP}.
This lack of transparency creates a critical awareness gap: users significantly underestimate the tracking permissions they grant~\cite{warin2025privacy} and the granularity of inferred information, such as emotional or cognitive states~\cite{hadan2024}, increasing their vulnerability to hyper-targeted influence, such as attention-driven targeted advertising~\cite{abraham2022}.

Large Language Models offer a promising approach to addressing this gap. 
Conversational agents can summarize complex data practices, adapt explanations to user literacy levels, and answer situational questions~\cite{prismeStudy,sun2025empowering}.
These capabilities could address key design challenges for privacy-related user interfaces in VR, such as balancing user engagement with privacy awareness and breaking down privacy policy information for user comprehension~\cite{xrAwareness, paneva2024ieeepvc}.

To explore this potential, we embed an LLM-powered privacy assistant directly at the point of decision -- a VR app store. The assistant is presented in two interaction modes: a text-based floating chat panel and an embodied virtual avatar providing spoken explanations.
Our work is guided by the following research questions:
\begin{enumerate}
\item[\textbf{RQ.1}] How can LLM-powered privacy assistants impact privacy awareness and decision-making in VR?

\item[\textbf{RQ.2}] How does interface modality (text chat vs.\ avatar) affect user comprehension, interaction, and overall experience?
\end{enumerate}

We address these questions through an exploratory within-sub\-jects study (N=21), comparing unassisted browsing with the two assisted conditions: a text-based chat panel and an embodied avatar. 
In semi-structured interviews, we qualitatively investigate participants' decision-making, modality preferences, and tool interaction.

Our findings show that, both the avatar and chat conditions supported participants’ perceived privacy risk comprehension and awareness during app selection, with participants more frequently ruling out applications they perceived as higher risk.
The avatar supported a more natural and socially engaging interaction for some participants, while others experienced it as distracting or uncanny and preferred the text-based chat interface. The chat modality enabled careful reading and reflection, though some participants reported feeling overwhelmed by the volume of text.
Overall, our early results suggest that both embodied virtual agents and text-based chatbots present viable approaches for supporting informed consent in VR, while warranting larger-scale validation across diverse VR contexts and user populations.

Our \textbf{contributions} are twofold: 
1) \textbf{Design and implementation of an LLM-powered privacy assistant integrated into a VR app store}, demonstrating how privacy policy assistance can be embedded directly at the point of decision through both text-based and embodied interaction modalities.
2) \textbf{Empirical evaluation of chat-based and embodied privacy assistance}, providing qualitative insights into impact on privacy awareness and user experience, as well as the modality-specific trade-offs.%

\section{Related Work}
\label{sec:rel_work}

Prior work spans empirical risk assessments, transparency gaps, and automated privacy policy assistants.

\textbf{Biometric Risks.}
VR headsets expose users to privacy risks distinct from web and mobile ecosystems. Even though generalizability across applications is contested by some research~\cite{schach2025motion,wang2024cross}, continuous telemetry streams create unique biometric fingerprints: motion trajectories from head and hand sensors can identify individuals within seconds with near-perfect accuracy across sessions~\cite{privacyRisks,motionIdentification,motionIdentification50k}. Similarly, eye-tracking data has been shown to effectively identify individuals~\cite{David-John} and reveal cognitive load and emotional arousal~\cite{eyeTracking}. Biometric data can reveal sensitive health indicators or personality traits~\cite{abraham2022}. 
However, users are largely unaware to the extent of these capabilities, often assuming only physical movement is tracked, overlooking deeper emotional or cognitive inferences~\cite{hadan2024}. Moreover, biometric signatures derived from VR interaction remain stable over months, rendering traditional anonymization ineffective~\cite{XR}.
Despite this heightened exposure, transparency mechanisms in VR remain inadequate. Privacy disclosures are frequently incomplete~\cite{Zhan2024VPVetVP,XR} and often buried in nested menus that are only accessible after installation~\cite{xrAwareness}.
As a result, users are frequently uninformed about data practices~\cite{abraham2022}.%

\textbf{AI-Driven Privacy Assistants.}
VR environments require mechanisms that make abstract data risks visible and understandable to users~\cite{david2021let}, yet few assistants operate in VR contexts or prioritize user trust~\cite{cornellLitReview}.
Beyond VR, research has long sought to automate the analysis of privacy policies. 
Early approaches leveraged classical natural language processing (NLP) and datasets such as OPP-115~\cite{wilson2016creation} to classify policy text. 
Tools like \textit{Polisis}~\cite{harkous2018polisis} and \textit{PriBot}~\cite{harkous2016pribots} pioneered this space by generating icon-based ratings or retrieving policy segments in response to chat queries. Nevertheless, these systems typically returned raw policy excerpts, leaving the burden of legal interpretation on the user.

Subsequent tools attempted to improve contextuality and usability.
\textit{PrivacyInjector}~\cite{windl2022automating} overlaid icons directly onto website elements, while \textit{PrivacyCheck}~\cite{nokhbeh2020privacycheck} scored policies against predefined criteria (e.g., GDPR compliance). Although such interventions supported more informed decision-making, they remained largely rigid, as users could not query specific concerns or request simplified explanations for complex data flows.

Recent work has increasingly turned to LLMs to bridge this comprehension gap. Studies show that LLMs match or exceed traditional NLP approaches in extracting data types~\cite{rodriguez2024large,zhang2025privcaptcha} and can reduce cognitive load by summarizing risks~\cite{chen2025clear,sun2025empowering}. 
For instance, \citeauthor{sun2025empowering} demonstrated that an LLM-based agent can effectively categorize and summarize policy sections, increasing user confidence~\cite{sun2025empowering}.  
Similarly, the PRISMe browser extension showed that interactive, LLM-based exploration privacy policies can improve policy understanding and risk awareness during browsing~\cite{prismeStudy}.

\textbf{Summary and Research Gap.} 
While VR systems enable granular and persistent data collection, current transparency mechanisms are insufficient and poorly integrated.
While LLM-based privacy assistants show promise for improving policy comprehension, they lack integration into immersive, point-of-decision contexts. %
It remains unclear how such assistants should be integrated into VR environments or how interaction modalities affect user comprehension, usability, and privacy-related decision-making.
To address these gaps, we embed an LLM-powered privacy assistant into a VR app store and 
compare text-based versus embodied modalities for privacy-aware VR app selection.

\section{System Design}\label{sec:sys}
\begin{figure}[t]
  \centering
  \begin{subfigure}[b]{0.475\textwidth}
    \includegraphics[width=\textwidth]{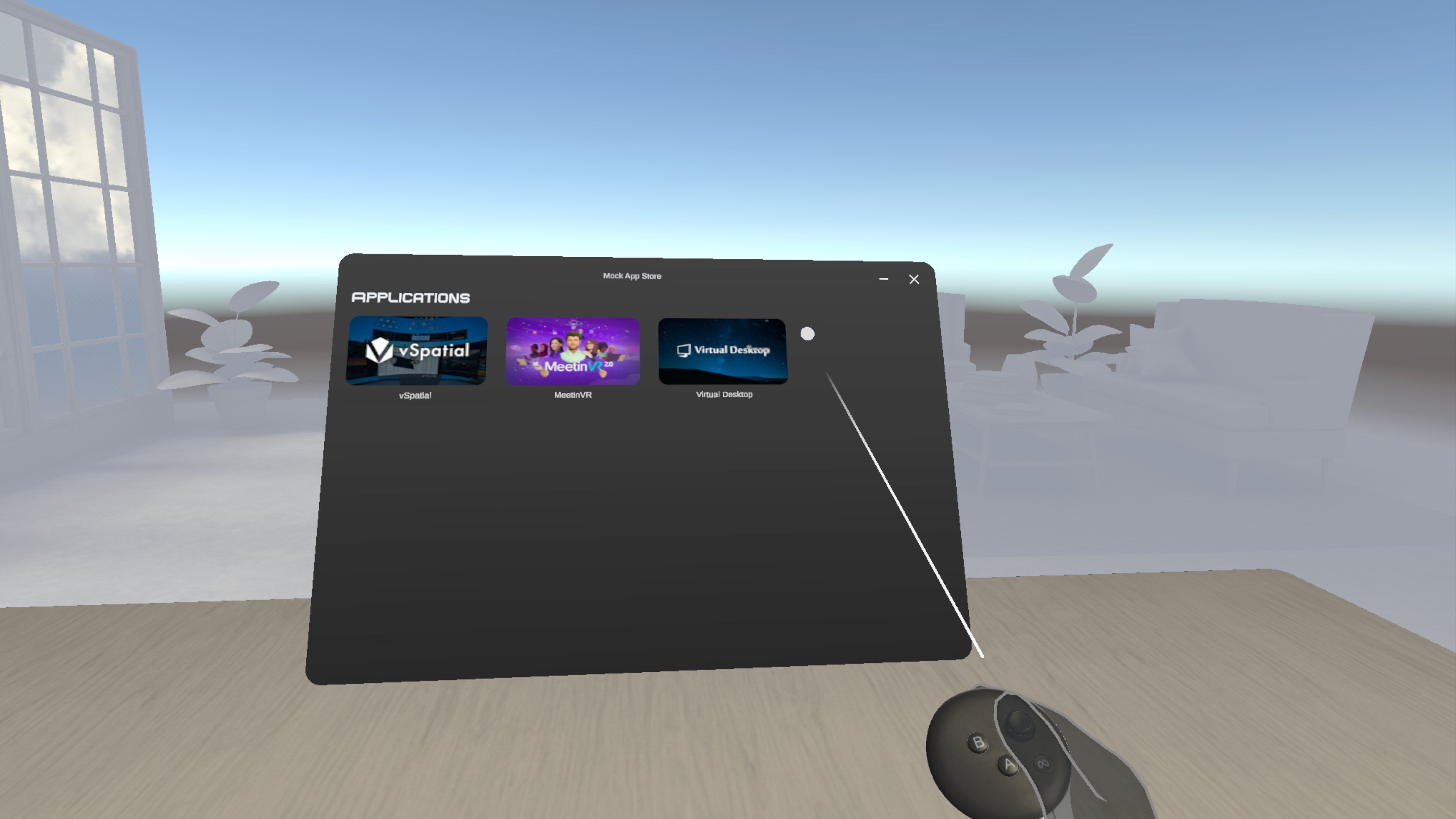}
    \subcaption{Initial VR app store scene}
    \label{fig:initialScene}
  \end{subfigure}
  \hfill
  \begin{subfigure}[b]{0.475\textwidth}
    \includegraphics[width=\textwidth]{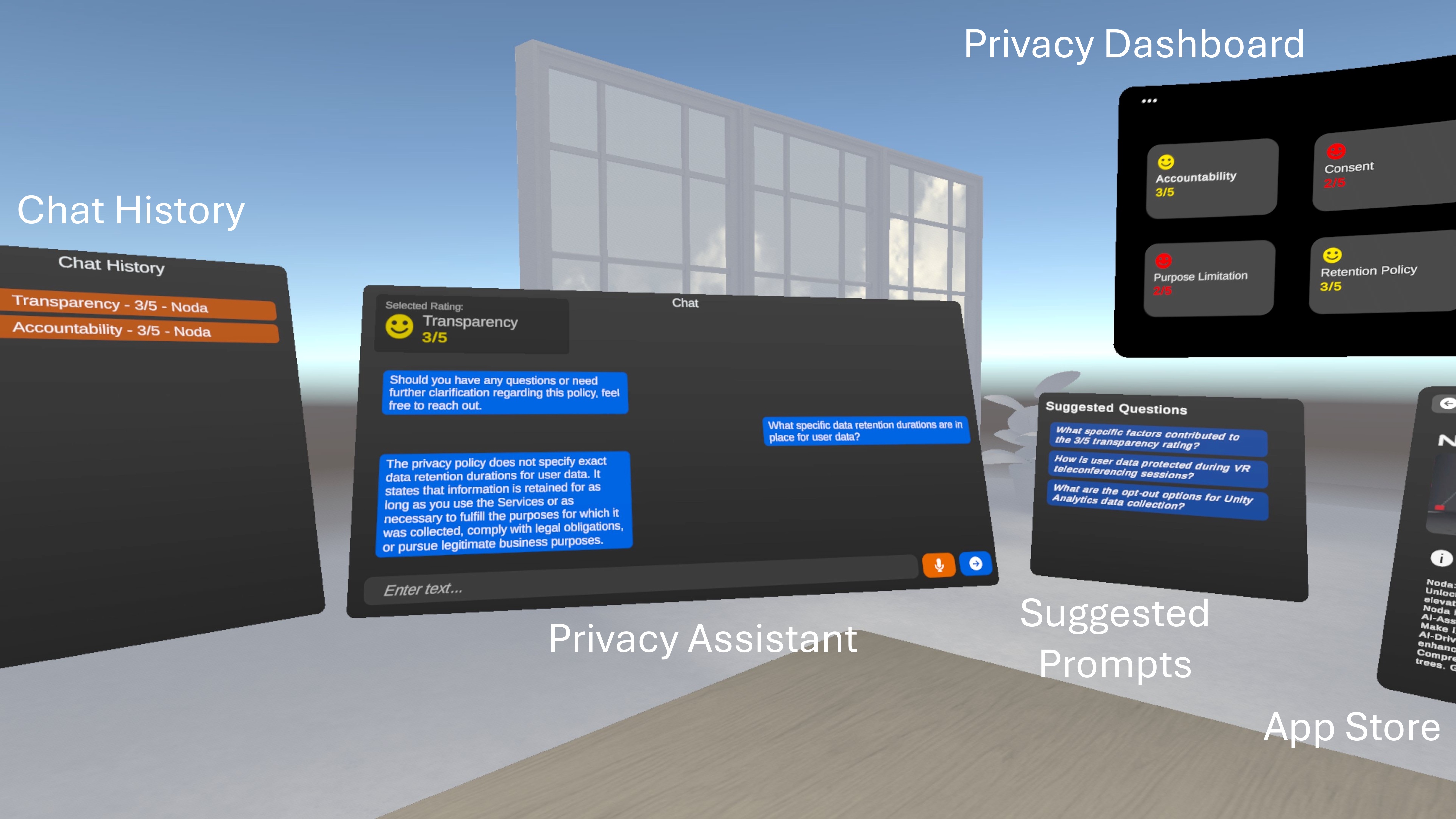}
    \subcaption{Chat-based interaction}
    \label{fig:img2}
  \end{subfigure}
  \hfill
  \begin{subfigure}[b]{0.475\textwidth}
    \includegraphics[width=\textwidth]{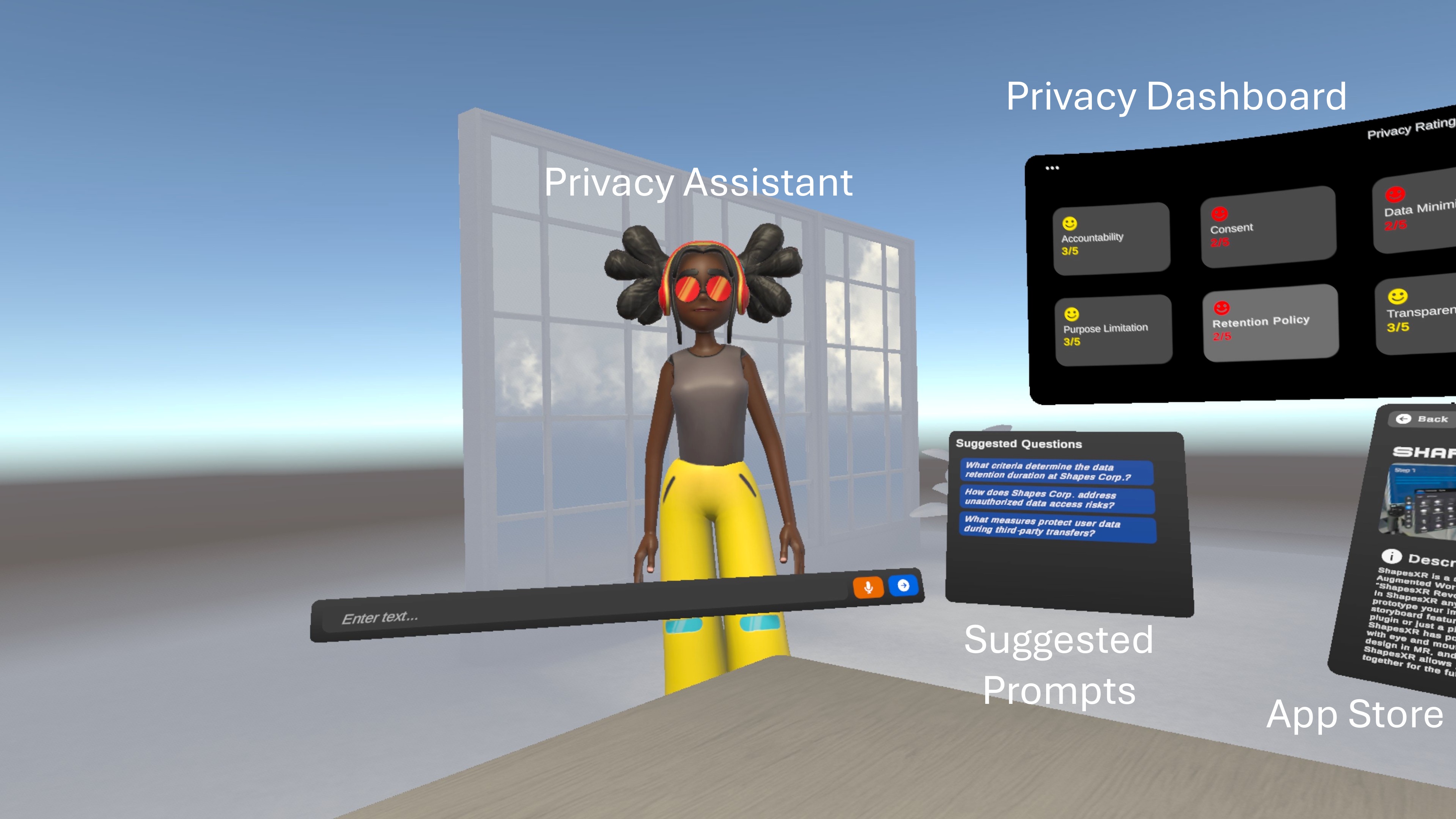}
    \subcaption{Avatar-based interaction}
    \label{fig:img3}
  \end{subfigure}
  \caption[Illustration of all VR-scenes]{%
    Illustration of the three VR interface conditions used in the user study.
  (a) \textit{Baseline: }storefront-only app store view (b) \textit{Chat condition: } text-based interface and privacy ratings. (c) \textit{Avatar condition: } spoken explanations, gesture cues and privacy ratings.}
  \label{fig:three_panel}
  \vspace{-0.5cm}
  \Description{Illustration of three virtual reality app store interfaces used in a user study. The top panel shows a simple VR environment with a floating rectangular display that lists several apps on a dark storefront screen, viewed from the perspective of a seated user holding a motion controller. The middle panel shows the same VR room with multiple floating panels including a central chat window displaying blue text message bubbles, side panels with chat history and suggested questions, and an app information area, representing a text-based chat interaction for exploring privacy information. The bottom panel shows the room again, now with a stylized humanoid avatar standing in front of the user; additional floating panels to the right present privacy policy information and ratings, exemplifying an avatar-led, speech and gesture–based explanation of app privacy in the VR store.}
\end{figure}

We developed a VR application in Unity that simulates an app marketplace, running on a Meta Quest 3 headset.  The user is positioned at a virtual desk with a floating, repositionable UI panel in front of them. Selecting an app opens a product card that mirrors standard Quest 3 storefronts, displaying images, description, and pricing information. 
To minimise confounding effects arising from app genre, each experimental condition included different, existing applications sampled from the same productivity-focused categories (e.g., virtual meetings, virtual desktop, or 3D sketching; see Appendix~\ref{app_apps}).
The privacy assistant employs a layered design approach inspired by PRISMe~\cite{prismeStudy}, guiding users from a privacy dashboard to either a chat or an avatar to query about specifics, depending on the experimental condition.
Here, both interfaces display a set of three context-aware suggestions to support users with limited privacy expertise.
The dashboard displays ten "traffic-light" (red-yellow-green) indicators representing specific data categories (e.g., Sensor Access, User Rights). This allows users to identify specific risk factors at a glance.

Users are presented with the following interface variations:

\textbf{Baseline (Storefront Only):} The control condition displays only the standard product card (Figure~\ref{fig:initialScene}) and descriptions, without the privacy assistant. %

\textbf{Chat Assistant:} 
In this condition, a privacy dashboard appears left of the main store window (Figure~\ref{fig:img2} upper right). Users can select a rating category to receive a text summary or type/speak free-form questions in a chat panel (Figure~\ref{fig:img2} middle). A scrollable transcript (Figure~\ref{fig:img2} left) maintains conversation history for later review.

\textbf{Avatar Assistant:} 
In this condition, instead of a chat window, a 3D humanoid character (sourced from Mixamo\footnote{www.mixamo.com}) 
appears behind the desk (Figure~\ref{fig:img3}). The avatar utilizes uLipSync~\cite{ulipsync} for real-time lip-synchronization and procedural gestures. It responds to the same inputs as the chat interface but delivers information via synthesized speech and non-verbal cues.

Both chat and avatar conditions share the same LLM backend and visual design, differing only in their presentation modality.

\textbf{Backend.}
\label{sec:sysarc}
For each app, the corresponding privacy policy text was cached locally and provided to the LLM at query time.
We utilize OpenAI's GPT-4o API~\cite{OpenAI2024} for dashboard data generation and question answering functionality. 
We borrowed and extended the prompting for generating ratings from PRISMe~\cite{prismeStudy} and adjusted chat prompting to our VR context (see Appendix~\ref{prompt}).
Voice interaction is handled via Meta’s Wit.ai service~\cite{witai2024} for speech-to-text and text-to-speech conversion.

\section{User Study}
We conducted a within-subjects lab study to investigate how different privacy interface conditions influence users’ privacy awareness understanding, and overall experience. 
In the \textit{Baseline} condition, participants viewed static app descriptions and standard storefront information without interactive privacy support. The \textit{Chat} and \textit{Avatar} conditions each introduced an integrated LLM-powered privacy policy assistant.

\textbf{Participants.}
We recruited 21 participants (9 female, 12 male). The sample included 14 students and 7 employed professionals. Sixteen participants had prior VR experience (primarily gaming), while five were novices. The study lasted approximately 60 minutes, and participants received either €12 or course credits as compensation.

\textbf{Procedure.} After providing informed consent, participants completed a demographic questionnaire, the Affinity for Technology Interaction (ATI) scale~\cite{franke2019personal}, and the Internet Users’ Information Privacy Concerns (IUIPC) scale~\cite{malhotra2004internet} to establish technical affinity and baseline privacy attitudes.

Participants then completed three VR sessions.
The first session was always the \textit{Baseline} 
to capture normative unprimed app-browsing behavior, 
followed by the \textit{Chat} and the \textit{Avatar} conditions in a counterbalanced order.
In each session, participants were presented with three productivity apps (different apps in each condition). They were instructed to browse the app store, think aloud, and select one app they would pick to "install on your own device" or reject all options.
In both assistant conditions, selecting a privacy rating triggered an auto-generated overview. Participants could then ask unstructured questions via voice input, text input, or suggested prompts until they felt ready to make a decision.

After completing all three VR sessions, participants took part in a semi-structured interviews focusing on their interaction experience, decision making process, and modality preferences (see Appendix~\ref{app_interview} for the interview guide). %

\textbf{Data Analysis.} Interview transcripts were analyzed using inductive thematic analysis~\cite{blandford2016qualitative}. Two researchers independently coded the data using Atlas.ti~\cite{atlasti} and then met to discuss discrepancies and iteratively refine the codebook, resulting in 16 codegroups (see Supplementary Material). %

\section{Results}

To contextualize our qualitative findings, we report participants' technical affinity and baseline privacy attitudes. 
Scores on the ATI scale indicated high technical affinity in general ($M=4.20, SD=0.80$).
Participants also reported relatively high privacy concern across the IUIPC dimensions of \textit{Awareness} ($M=5.95, SD=1.31$), \textit{Control} ($M=5.67, SD=1.28$), and \textit{Collection} ($M=5.38, SD=1.89$). 

All participants inspected for each app at least two rating categories
and posed an average of 9
follow-up questions. %
Three categories accounted for 53\% of all look-ups: \textit{User Rights} (22\%), \textit{Consent} (17\%), \textit{Sensor Access} (14\%).
Next, we present our qualitative results, structured around the four identified main themes.

\textbf{Balancing Engagement and Control.}
Some participants found the avatar's presence engaging and natural (P10, P12, P15), while others experienced the anthropomorphism as uncomfortable -- for example, P6 \textit{"hated having someone stare at [them]"} -- or disliked adapting to its fixed speech pace (P8, P13-14). 
Participants also reported several usability frictions, including the system's inability to handle natural speech pauses, causing interruptions (P7), having to click on UI elements disrupting conversation flow (P1), and a
lack of option to read responses alongside the audio (P2, P6, P7, P21) or review conversation history (P1, P5, P7, P9).
The chatbot addressed some of these limitations by supporting review and reflection (P1-2, P5, P7-9), allowing users to \textit{"thoroughly understand and check the content of the answer provided"} (P2).
Overall, participants praised the tool's robustness to spelling or pronunciation errors and reliable intent recognition (P1).
The tool's intuitive design (P1, P7, P9, P16, P18), being \textit{"easy to understand and [...] feel[ing] like it is for anybody"} (P11) helped users. %
Overall, balancing engagement and control emerges as a key design challenge, with embodied agents drawing users in while textual interfaces better support deliberate reflection.

\textbf{Layered Privacy Information and Personalization Needs.}
Participants generally appreciated the layered design with the dashboard providing a \textit{"one view shot"} overview (P1), and deeper inquiry options facilitated through follow-up questions, which helped users' understanding (P1-2, P10, P21). 
However, preferences varied: some would have preferred directly posing questions without first navigating the dashboard (P1), while others asked for more detailed justifications for individual ratings upfront (P5, P16, P19-20), including policy excerpts or concrete examples (P8, P16). 
Perceptions of information density also differed. While some found the amount of information to be \textit{"at a good level"} (P4) or valued them for providing meaningful choice (P10), others experienced the ten dashboard criteria as overwhelming (P5-6).
These divergent reactions highlight the need for personalization, including adjustable information granularity and presentation.
Additional suggestions included an at-a-glance overall privacy score for easier app comparison (P2, P4, P6, P21), customizable panel layouts (P5, P9, P19), adjustable UI sizing (P14, P19), and configurable VR scenes/backgrounds (P7, P12, P21).
Participants emphasized that such a tool would only have meaningful impact if seamlessly integrated into the app store experience like in our prototype (P7, P20).

\textbf{Raising Privacy Awareness and Understanding.}
Prior to using the tool, many participants reported indifference towards privacy policy information and described routinely accepting tracking without scrutiny (P1, P7, P9, P14).
Interaction with the privacy assistant prompted several users to shift from indifference to a more active evaluation of data practices.
For example, P14 realized, \textit{"what they [the app] were tracking and some things I might not have been comfortable with"}. 
The dashboard's quick overview, combined with question-answering functionality that \textit{"quite consistently added information"} (P2) supported \textit{"informed decision[s]"} (P2), and encouraged participants to consider privacy more deliberately in their decision-making. P1 reflected that, \textit{"we all need to be more cautious, and this panel really helps"}.
Several participants further reported intentions to engage more critically with privacy policies in the future, even in the absence of a dedicated privacy policy assessment tool (P1-2, P14). %
At the same time, one participant reflected on issues of trust in AI-generated explanations, noting the importance of maintaining a critical stance and indicating they would \textit{"read the actual privacy policies and compare them against the chatbot results"} (P6). %

\textbf{Privacy as a Veto Mechanism in App Selection.}
Participants rarely described privacy as the primary driver for selecting a VR application (P6, P10, P20). 
Instead, it functioned as a veto mechanism %
(P6, P10-11, P19). 
Across conditions, participants prioritized pragmatic and hedonic factors, such as price (P1, P5, P8, P16), app-specific utility (P7-10, P14, P20), and visual appeal (P9, P12-13, P15). 
Participants described the privacy assistant as particularly useful for assessing apps that lacked \textit{"instinctive trust"} (P10), highlighting its role in trust calibration. Some participants used the question-answering functionality to probe specific concerns, such as the availability of opt-out options (P6).
Others reported eliminating options by ruling out apps that received poor privacy ratings on the dashboard (P7, P11, P13, P20).

\section{Discussion}
\label{sec:disc}

We discuss how LLM-powered privacy assistants can impact privacy awareness and decision-making in VR (RQ1) and how interface modality (text chat vs. avatar) affects user comprehension, interaction, and overall user experience (RQ2).

\textbf{Addressing RQ1: }
Our findings suggest that the LLM-powered privacy assistant functions primarily as a \textit{supporting decision scaffold} rather than as a primary decision factor. %
While visual appeal, price, and perceived functionality remained dominant factors, the assistant prompted users to engage more deliberately with privacy-related information and reflect on the associated risks.
By embedding privacy support directly into the VR app store and surfacing risks at the point of decision, the assistant reflects core Privacy by Design principles of proactive disclosure and privacy embedded into system design~\cite{cavoukian2009privacy}.
While users valued the assistant’s granular breakdown of privacy criteria, the presence of multiple categories occasionally led to information overload, highlighting the potential value of progressive disclosure and personalization mechanisms that adapt the level of detail to users’ needs and momentary attention.
Finally, several participants emphasized that the assistant’s impact would be contingent on seamless integration into existing VR app store workflows, highlighting that the effectiveness of privacy assistance tools is shaped not only by their information quality, but also by frictionless embedding within users’ established interaction routines.

\textbf{Addressing RQ2: }
With respect to interface modality, our findings point to a trade-off between engagement and comprehension.
The embodied avatar often served as an effective entry point for initiating interest in privacy-related information, with some participants reporting higher engagement and trust, in line with prior work showing that anthropomorphic agents can foster perceived empathy and trust~\cite{ma2025effect}.
At the same time, other participants felt uncomfortable in the avatar’s presence or constrained by its fixed pacing.
In contrast, many participants preferred the text-based chat for understanding complex data practices, emphasizing the ability to re-read and process information at their own pace.
This suggests that spoken summaries may be well-suited for initial low-friction exploration, whereas deeper engagement benefits from reviewable, text-based information. 
Future work could explore hybrid interfaces that combine the avatar's engaging qualities with a textual transcript, allowing users to switch modalities depending on task complexity and preference.

\textbf{Limitations.}
Methodically, the lab setting and think-aloud protocol might have heightened participants' privacy awareness, potentially leading to more privacy-conscious behavior than in everyday use. 
Furthermore, the study focused on productivity applications, and decision-making may differ in social or gaming environments characterized by stronger immersion and social pressure, which might attenuate the impact of privacy cues.
Moreover, our sample of participants skewed towards younger, tech-literate participants.
Future work could focus on a longitudinal study, include various application genres, and look into whether and how various populations experience different levels of engagement or friction. 

While the Baseline condition was always presented first to capture unprimed browsing behavior prior to exposure to privacy assistance, this fixed ordering may have introduced learning or sensitization effects.
Future work should therefore randomize or temporally separate baseline exposure to better control for potential carryover effects.
Finally, future studies could incorporate objective comprehension and awareness measures to complement self-reported insights this study focused on.

As with AI-driven assistance more broadly, reliance on LLM-based explanations warrants caution, including the risk of hallucination. 
While this work focused on interaction modality rather than validating the LLM's output, mitigations involving Retrieval Augmented Generation (RAG) and Hallucination Aware Tuning have shown to be effective in reducing hallucinations~\cite{song-etal-2024-rag,freiberger2026helping} and could be implemented in future iterations.
Prior work on conversational XAI has shown that, while dialogue-based systems can improve understanding, they might also increase over-reliance on AI recommendations~\cite{overreliance}. 
This highlights the importance of designing privacy assistants that support critical engagement and trust calibration.

Our system relied on a remote LLM and a voice interaction service, %
meaning that user interactions may themselves be subject to external data processing. 
While this reflects common deployment practices, it introduces a trade-off between explanation quality and the privacy guarantees of the assistant itself.
Future work should explore this tension between user privacy (favoring smaller, locally deployed LLMs) and explanation quality (favoring larger, typically commercial models like GPT-4o, which we used).

\section{Conclusion}
We introduced an LLM-powered privacy assistant embedded in a VR app store and compared unassisted browsing with two assisted conditions, a chatbot and an embodied avatar, in a within-subjects study (N=21).
Our findings showed that both forms of assistance encouraged more deliberate engagement with privacy information, with privacy information primarily functioning as a veto mechanism to rule out applications with unacceptable data practices.
The embodied avatar fostered interest and social presence for some users, whereas the text-based chat interface better supported careful review and self-paced reflection.
This work demonstrates the potential of AI-driven privacy assistants to make privacy considerations more accessible and actionable at the point of decision in virtual immersive environments.

\begin{acks}
This work has received funding from the German Research Foundation (DFG) under grant agreement no. 521584224.
Vincent Freiberger and Arthur Fleig acknowledge the financial support by the Federal Ministry of Research, Technology and Space of Germany and by Sächsische Staatsministerium für Wissenschaft, Kultur und Tourismus in the programme Center of Excellence for AI-research „Center for Scalable Data Analytics and Artificial Intelligence Dresden/Leipzig“, project identification number: ScaDS.AI.
\end{acks}
\bibliographystyle{ACM-Reference-Format}
\bibliography{bibliography}
\appendix

\section{Applications Used in the Study}
\label{app_apps}

The following VR productivity applications were used across the experimental conditions.

\begin{itemize}
    \item Immersed (\url{https://immersed.com})
    \item Meta Horizon Workrooms (\url{https://forwork.meta.com/de/horizon-workrooms})
    \item MeetinVR (\url{https://www.meetinvr.com})
    \item Virtual Desktop (\url{https://www.vrdesktop.net})
    \item vSpatial (\url{https://www.vspatial.com})
    \item Noda (\url{https://noda.io})
    \item Gravity Sketch (\url{https://gravitysketch.com})
    \item ShapesXR (\url{https://www.shapesxr.com})
    \item Figmin XR (\url{https://overlaymr.com})
\end{itemize}

\section{Prompting}
\label{prompt}
\subsection{Dashboard Data}
    \textbf{System Prompt: }
    Your output must be a maximum of 600 words long! You are an expert in data protection and a member of an
ethics council. You are given a privacy policy. Your task is to uncover aspects in data protection declarations that
are ethically questionable from your perspective. Proceed step by step:
\begin{enumerate}
    \item Criteria: From your perspective, identify relevant ethical test criteria for this privacy policy as criteria
for a later evaluation. When naming the test criteria, stick to standardized terms and concepts that are
common in the field of ethics. Keep it short!
\item Analysis: Based on this, check for ethical problems or ethically questionable circumstances in the privacy
policy.
\item Evaluation: Only after you have completed step 2: Rate the privacy policy based on your analysis regarding
each of your criteria on a 5-point Likert scale. Explain what this rating means. Explain what the ideal case
with 5 points and the worst case with one point would look like. The output in this step should look like
this:

[Insert rating criterion here]: [insert rating here]/5 [insert line break]
[insert justification here]
\item Contextualization: 
Use the **attached images and the description of the VR application** to establish a contextual link between the privacy policy and the actual use case of the VR game or experience. Assess whether the data protection measures described in the policy appropriately address the specific risks, user interactions, and immersive data environments involved in the application. Take into account unique concerns such as biometric tracking, gaze monitoring, spatial mapping, and behavioral profiling that are often present in immersive technologies.
\item Conclusion: Reflect on your evaluation and check whether it is complete.
\end{enumerate}
\noindent
Important: Check for errors in your analysis and correct them if necessary before the evaluation. You must
present your approach clearly and concisely and follow the steps mentioned. Your output must not exceed 600
words.

    \noindent
    \textbf{User Prompt:} <Privacy Policy inserted here>
    
\subsection{Prompting Approach for Chat Interface}
\label{sec:chat_prompt}

\textbf{System Prompt: }Privacy policy: <Privacy Policy inserted here> | Rating: <Rating category inserted here> with rating <Rating inserted here>

\noindent
Users want to know more about how this rating is justified in the privacy policy. When answering the questions, focus on the given topic of the rating. Keep it short!

\noindent
\textbf{User Prompt: }<User Question inserted here>

\subsection{Prompting Approach for Avatar}
\label{sec:avatar_prompt}

\textbf{System Prompt: }Privacy policy: <Privacy Policy inserted here> | Rating: <Rating category inserted here> with rating <Rating inserted here>

\noindent
Users want to know more about how this rating is justified in the privacy policy. When answering the questions, focus on the given topic of the rating. Make sure that the answer is MAX. 25 WORDS! Keep the answer in a conversational style. The output will be transmitted to a TTS interface! Make sure to make full sentences which are understandable!

\noindent
\textbf{User Prompt: }<User Question inserted here>

\subsection{Prompting Approach for our Suggested Question Generation}
\label{sec:suggestion_prompt}

\textbf{System Prompt: }Your task is to generate thoughtful and critical questions regarding a privacy policy document. These questions should aim to clarify the scope, data handling practices, user rights, or compliance aspects of the policy. Your output must consist of exactly three well-structured questions. Format your output as valid JSON, using the following structure: { "suggestions": [ "Question 1", "Question 2", "Question 3" ] } Please ensure the questions are precise, relevant, and helpful for evaluating the transparency and completeness of a privacy policy. Make sure the questions are MAX. 9-14 WORDS!

\noindent
\textbf{User Prompt: }Specifically: Generate questions about the privacy policy of the app <App Title>, which has a rating of <Rating category inserted here>  with a score of <Rating inserted here>. Focus particularly on aspects that led to this rating. Consider critically examining any concerning practices or noteworthy protections mentioned in the following policy: <Privacy Policy inserted here>

\section {Interview Guide}
\label{app_interview}

Overall Experience
\begin{itemize}
\item Describe in your own words your experience with the two privacy policy interfaces (Chat and Avatar).  
\end{itemize}

\noindent Interface Comparison
\begin{itemize}
\item What are some positive and negative aspects of each interface in your own opinion?
\item How would you compare the interaction experience with each? 
\item How well were you able to obtain the information you wanted to know with each one?
\end{itemize}

\noindent Decision Making
\begin{itemize}
\item What was your thought process or strategy for choosing the VR apps? Did the approach change between sessions?  
\item Do you think the privacy assessment tool influenced your decisions? If yes, how?
\item How might this experience impact your future decisions when it comes to choosing apps or reviewing privacy policies?
\end{itemize}

\noindent Improvements and Final Reflections
\begin{itemize}
\item How would you improve the privacy policy tool or its interface? Are there any features you wish would have been available when browsing the apps? 
\item Do you have any final thoughts or feedback about your experience in the study? 
\end{itemize}

\section*{Supplementary Material: Codebook}

\label{codebook}
In the following we report the full codebook resulting from thematic analysis of the semi-structured interview for transparency.

\begin{table*}[hb]
\caption{Codebook resulting from thematic analysis of the semi-structured interviews}
\small
\begin{tabular}{|p{5 cm}|p{12 cm}|}
\hline
\textbf{Code}                                                   & \textbf{Description}                                                                   \\

                    \hline

\multicolumn{2}{|p{14.3 cm}|}{\textbf{Codegroup 1:} Avatar Negative}\\
\hline

difficult to fully grasp information             &    Just spoken and not written presentation of information makes it difficult to fully grasp the presented information                                     \\ \hline

have to adapt to avatar speaking pace                 &       Avatar has a fixed voice output users need to adapt to                       \\ \hline

interruptions / fluid conversation expected                                   &    Smaller speech pauses while formulating the question cause avatar to interrupt user who is not finished asking the question            \\ \hline

less efficient                                    &    Constrained to the duration of the audio at the given talking speed causes perception of slow information provision           \\ \hline

not possible to read                    &      Users complain that they could not read the response          \\ \hline

presence of avatar disliked                               &      Some users felt uncomfortable talking to the digital avatar           \\ \hline

user prefers speech interaction only while driving                                       &       Speech interaction only preferred while driving       \\ \hline

\multicolumn{2}{|p{14.3 cm}|}{\textbf{Codegroup 2:} Avatar Positive}\\
\hline

comfortable                            &      Users like that they don't have to read a lot of text and can just listen          \\ \hline

good to avoid information overload                                   & Avoids reading and not understanding what you read                  \\ \hline

increased presence appreciated                               & Users like that talking to the avatar feels more like talking to a human                 \\ 
 \hline

more accessible (less tech savy, dyslectic,...)           & Some users feel like the avatar is simpler and more accessible to a wide range of users, particularly to reading impaired users                   \\ \hline

more interesting/engaging            & Users feel more engaged by interacting with the avatar and more interested in the interaction                    \\ \hline

spatial audio                 & Users like the three-dimensional feel of the audio of the avatar                  \\ \hline

visual and auditory aspects help clearer understanding                                     & Users feel like gestures and visual presentation of the avatar combined with the audio of its speech support comprehension                   \\ \hline

\multicolumn{2}{|p{14.3 cm}|}{\textbf{Codegroup 3:} Chat Negative}\\
\hline

dyslectics may struggle                           & Textual presentation of information may cause problems for dyslectics                  \\ \hline

nothing new and stimulating                                             & Users feel not stimulated or engaged enough with textual presentation                           \\ \hline

too much to read through                              & Users feel overwhelmed by the amount of text to read through                   \\ 
\hline
\multicolumn{2}{|p{14.3 cm}|}{\textbf{Codegroup 4:} Chat Positive}\\
\hline
better able to thoroughly understand                             & The written chat allows users to get a deeper, more thorough understanding of the presented information                                                                                                               \\ \hline

chat intuitive and easy to ask questions                    & For some users it felt esier and more intuitive to ask questions in the chat                                                                                                             \\ \hline

easy to understand for anybody                                & The chat provides simple language answers                                                                                                                                      \\ 
\hline

liked feeling anonymous to ask questions                                        & Feeling more anonymous made it easier to ask questions                                                                                                   \\ \hline

history feature was helpful                                    & Seeing the chat history was perceived helpful by users      \\                                                                                                  
\hline

liked the speech-to-text input                                    & Speech-to-text was liked to input questions      \\                                                                                                  
\hline

preference for chat, because it is faster than avatar                                    & Some users felt they could grasp important content quicker by reading through it with the chatbot      \\                                                                                                  
\hline

question selection worked well                                    & Selecting suggested questions was appreciated and worked well      \\                                                                                                  
\hline

read at own pace                                    & Users liked that they could read through chat responses at their own pace      \\                                                                                                  
\hline

suggested questions helped gain a more complete understanding                                   & Using suggested questions helped users build a more complete understanding of privacy implications      \\                                                                                                  
\hline

summarization liked                                    &   Users like that the chat summarizes information well    \\                                                                                                  
\hline

user likes to read, so prefers the chat interface                                    &    Some users generally prefer reading over listening  \\                                                                                                  
\hline

users can better express themselves via text                                    &   Some users find it easier to phrase questions via text \\                                                                                                  
\hline

\end{tabular}
\end{table*}

\begin{table*}[hb]
\small
\begin{tabular}{|p{5 cm}|p{12 cm}|}
\hline
\textbf{Code}                                                   & \textbf{Description}                                                                   \\
\hline

\multicolumn{2}{|p{14.3 cm}|}{\textbf{Codegroup 5:} Decision Factors}\\
\hline

coolness factor                                    & VR app selected based on how cool it was perceived to be      \\                                                                                                  
\hline

decision based on best functional fit with personal interests                                    & VR app selected based on how well its functionality fits users individual interests      \\                                                                                                  
\hline

most privacy protection                                    & Users chose the VR app which was the most protective of their privacy      \\                                                                                                  
\hline

price                             & Price as the decisive decision driver      \\                                                                                                  

\hline

ability to opt-out                           & opt-out option as a prerequisite for accepting a vr app      \\    
                    \hline

privacy as secondary decision factor                            & privacy was considered as a secondary factor in the decision process      \\                                                                                                  
\hline

simplicity combined with functionality                            & users liked vr apps that kept it simple while providing the requested functionality      \\                                                                                                  
\hline

visually pleasing                            & Users chose VR apps as they thought their design was visually pleasing      \\                                                                                                  
\hline
\multicolumn{2}{|p{14.3 cm}|}{\textbf{Codegroup 6:} Decision Process}\\
\hline

few dashboard criteria investigated                            & Users chose only a few of the dashboard criteria to investigate and make up their mind for the decision      \\                                                                                                  
\hline

privacy information to "verify" decision                           & Users use the privacy information to confirm or invalidate the app they want to decide on      \\                                                                                                  
\hline

regarding privacy only data collection is deemed relevant                            & Users only considered data collection information when investigating privacy     \\                                                                                                  
\hline

same strategy for choosing apps for both interactions                            & Users did not adapt their decision making process given the privacy assistant      \\                                                                                                  
\hline

step by step iteratively eliminating options ending up with privacy as a decision driver                            & Users iteratively eliminate options with privacy used in the last elimination step as a decision driver      \\                                                                                                  
\hline

user chose the most useful app with decent privacy rating                            & Users the most useful app out of all apps with decent privacy rating by the tool      \\ 

\hline

user preferred the apps that were transparent about their data processing                            & Some users chose only apps that were transparent about their data processing      \\ 

\hline

user ruled out the app with bad privacy scores, regardless of the functionality                            & Users ruled out apps with bad privacy scores, regardless of how much they liked their functionality      \\ 

\hline
\multicolumn{2}{|p{14.3 cm}|}{\textbf{Codegroup 7:} Decision Influence by Tool}\\
\hline

taking into account visual smiley cues                            & Users base part of their decision making on the visual smileys provided by the tool      \\ 

\hline

dashboard ratings impacted decision making                            & Users considered the ratings on the dashboard in their decision-making \\ 

\hline

didn't influence decision                            & Users stating they weren't influenced in their decision-making by the tool at all      \\ 
\hline

influenced decision                           & Users stating they were influenced in their decision-making by the tool\\ 

\hline

some impact where apps tracked more than comfortable with                            & When Users found out with the tool that VR apps collected more than they were comfortable with, the tool did impact their decision      \\ 

\hline

visual cues combined with in-depth questions effective decision support                            & Combining the smileys as visual cues with the in-depth exploration with questions provides good decision support      \\ 
\hline
\multicolumn{2}{|p{14.3 cm}|}{\textbf{Codegroup 8:} Future Decisions}\\
\hline

app utility focus                            & For future decisions, users consider app utility to be their deciding factor      \\ 

\hline

easily digestible format helps decision making                            & For future decisions, users consider involving privacy information if presented in an easily digestible format like the presented tool      \\ 

\hline

helps trust calibration                            & Users consider that for future decisions they might calibrate their level of trust with a tool like the one presented in the study      \\ 
\hline

low decision impact                            & Users like a privacy assistant as the one in the study would have a low decision impact on them in the future      \\ 

\hline

potential with increased relevance of VR                            & Users consider the rising relevance of VR to be a driver of the relevance of such a tool      \\ 
\hline

would take into account app privacy                            & Users say they will generally take into account privacy in their future decision-making      \\ 
\hline

would use if integrated in app store                           & Users state they would be impacted by a privacy assistant if integrated directly into the VR app store      \\ 
\hline
\multicolumn{2}{|p{14.3 cm}|}{\textbf{Codegroup 9:} Raised Awareness}\\
\hline

interest in use for potential of raised awareness and learning                            & Users state interest in the tool for the educational potential and raising their awareness      \\ 
\hline

raised awareness by tool usage                            & Users display raised awareness after using the tool      \\ 
\hline

low, nothing new learned                            & Some users report no effect on their awareness as they have learned nothing new      \\ 
\hline
\multicolumn{2}{|p{14.3 cm}|}{\textbf{Codegroup 10:} General Privacy Behavior}\\
\hline

accept without reading                            & Users accept all without considering the privacy policy information      \\ 
\hline

basic privacy behaviors                            & Users show basic privacy protective behaviors like declining cookies      \\ 
\hline

blindly accept privacy terms willingly                           & Users blindly and willingly accept privacy terms      \\ 
\hline

more careful about kids data                            & More protective about data of kids then own data      \\ 
\hline

privacy policy information not perceived as relevant to user                            & User does not consider privacy policy information as relevant      \\ 
\hline

\end{tabular}
\end{table*}

\begin{table*}[hb]
\small
\begin{tabular}{|p{5 cm}|p{12 cm}|}
\hline
\textbf{Code}                                                   & \textbf{Description}                                                                   \\
\hline

\multicolumn{2}{|p{14.3 cm}|}{\textbf{Codegroup 11:} VR Interaction}\\
\hline

initial overwhelm                            & Initial overwhelm caused by VR environment and interaction modality      \\ 
\hline

tedious to enter questions in VR                            & Typing in questions in VR is difficult      \\ 
\hline

trouble with controller                           & User faced difficulty with VR controller      \\ 
\hline

VR interaction draining                          & Users perceived VR as interaction modality to be draining      \\ 
\hline
\multicolumn{2}{|p{14.3 cm}|}{\textbf{Codegroup 12:} UI Improvements}\\
\hline

ability to organize windows                           & Users want to self-organize the position and sizing of the windows the tool displays      \\ 
\hline

adjustability UI element and font size for accessibility                            & Users suggest adjustability of UI elements to make the tool more accessible for everyone      \\ 
\hline

clunky UI                           & Some users perceive the UI to be clunky      \\ 
\hline

direct access to chat instead of via dashboard criterion                            & User wants to directly access the chat without having to go via the dashboard      \\ 
\hline

more interesting setting of VR environment                            & VR environment scene could be set more interestingly      \\ 
\hline

navigation issues                            & Users face issues navigating within the tool      \\ 
\hline

too many criteria on dashboard, information overload                            & Amount of criteria on the dashboard overwhelms users      \\ 
\hline

seperate windows unintuitive                            & Difficult to work with information displayed in multiple windows      \\ 
\hline

UI element sizing                            & Some UI elements were perceived to be inappropriately sized (buttons and icons larger)     \\ 
\hline

window management - one big window                           & Suggestion of one large window to display all information      \\ 
\hline
\multicolumn{2}{|p{14.3 cm}|}{\textbf{Codegroup 13:} Feature Improvements}\\
\hline

adjust avatar talking speed                            & Adjustable talking speed for the avatar      \\ 
\hline

adjustability volume of audio output for accessibility                            & Adjustability of the audio output volume for accessibility      \\ 
\hline

clickable chat history for avatar                            & Include a chat history for review for the avatar      \\ 
\hline

easier comparibility for multiple apps                            & Multiple apps should be made easier to directly compare      \\ 
\hline

last question feature                            & Make last question available to be asked again for different app     \\ 
\hline

improve explanation quality and simplicity                         & Provide better explanations which are yet simple to understand      \\ 
\hline

initial policy content overview / summary                        & Provide Instead of a dashboard present a short policy content overview / summary      \\ 
\hline

integrate feedback mechanism                         & Allow users to give feedback to the developers in the app      \\ 
\hline

lowest rating makes inquiry  unnecessary                         & User perceives the lowest rated criteria irrelevant to inquire      \\ 
\hline

More clarity and specificity in  explanations                        & Provide clearer and more specific explanations      \\ 
\hline
more context on evaluation, practical examples                         & Contextualize evaluations with practical examples for better understanding      \\ 
\hline

more differentiation between policy assessments needed                         & Policy assessments need to differentiate between apps more clearly      \\ 
\hline

more granular information on the dashboard, evaluation transparency                         & The dashboard needs explanations with more granular information which better explains the assessment results and gives evaluation transparency      \\ 
\hline

more sructured and condensed responses                         & Resonses should be more condensed and more structured e.g. with bullet points      \\ 
\hline

no overall score to compare the apps directly                         & There should be an overall score fore apps to facilitate comparison     \\ 
\hline

no, tool very good                         & No improvement suggestions, tool perceived to be good as it is      \\ 
\hline

too many criteria on dashboard, information overload                            & Amount of criteria on the dashboard overwhelms users      \\ 
\hline
unsure what to ask at times                         & User unsure about what to ask      \\

\hline

integrate privacy information into store app descriptions                          &  User would expect the privacy info to be in line with the app description     \\ 
\hline

\end{tabular}
\end{table*}

\begin{table*}[hb]
\small
\begin{tabular}{|p{5 cm}|p{12 cm}|}
\hline
\textbf{Code}                                                   & \textbf{Description}                                                                   \\
\hline

\multicolumn{2}{|p{14.3 cm}|}{\textbf{Codegroup 14:} Modality-agnostic Positive}\\
\hline

    ability to ask anything                     &   Users like that the tool allows them to ask any question on the privacy policy of the VR app    \\ 
\hline

    clear and well structured responses                     &   Users like that responses were clear and structured    \\ 
\hline

completeness                     &    The tool covers everything relevant regarding data protection for VR apps   \\ 
\hline

    correct intent recognition and robustness to typos or pronounciation                     &    The tool is not sensitive to typos or bad pronunciation and correctly identifies what the user wants to learn about \\ 
\hline

    digestible and helpful                     &    The information presented by the tool was helpful and easy to process   \\ 
\hline

    easy to use                     &    The tool did not require training, was self-explanatory, intuitive and easy to use   \\ 
\hline

    effective visual cues                     &    The smiley visualizations on the dashboard were effective in quickly communicating relevant judgement on the policy   \\ 
\hline

    good one-view shot overview                     &   The tool provided an effective overview to gather the most relevant information in one view    \\ 
\hline

    improved understanding                     &    The tool helped users improve their understanding of data protection issues   \\ 
\hline

    interactive presentation                     &   Users liked the interactive presentation of information    \\ 
\hline

    multiple windows liked, dragable                     &   Users liked that there were multiple windows and some were draggable    \\ 
\hline

    nice design                     &    Users like the overall design of the tool   \\ 
\hline

nice to have choice based on the categories                     &    Users like the choice the different criteria on the dashboard provide   \\ 
\hline

no typing necessary to ask questions                     &    Users like that they don't have to type and can also enter questions via voice input   \\ 
\hline

surrounding VR environment nice                     &    Some users like the VR environment scene the interaction was set in   \\ 
\hline

\multicolumn{2}{|p{14.3 cm}|}{\textbf{Codegroup 15:} Other Use Cases}\\
\hline

also for mobile app stores (Android/Apple)                         & Availability of such a tool in the Android or Apple mobile app stores would be desirable      \\ 
\hline

use case terms and conditions                         & Applying such a tool to terms and conditions would be desirable      \\ 
\hline

\multicolumn{2}{|p{14.3 cm}|}{\textbf{Codegroup 16:} Miscellaneous}\\
\hline

    creep out about AI capabilities                     &    Some users are generally creeped out about current AI capabilities   \\ 
\hline

    Decreasing use of chat functionality over time after added value educational                     &    The tool may first have educational value, but may be used less later   \\ 
\hline

    distrust toward AI                     &    Users have a general distrust toward AI and its reliability   \\ 
\hline

    stable network required                     &    Utilizing the tool requires a stable internet connection   \\ 
\hline

avatar and chatbot felt similar                                    & Users did not perceive avatar and chatbot too differently and could work with both      \\                                                                                                  
\hline

\end{tabular}
\end{table*}

\end{document}